\documentclass[12pt]{article}
\usepackage{times}
\usepackage[top=5cm,bottom=5cm,left=3cm,right=3cm]{geometry}                
\usepackage{graphicx}
\usepackage{amsmath}
\usepackage{enumitem}
\usepackage{amssymb}
\usepackage[english]{babel}
\usepackage[small,compact]{titlesec}
\usepackage{color}
\usepackage{tabu}
\usepackage{subcaption}
\usepackage{url}

\newcommand{\openr}{\hbox{${\rm I\kern-.2em R}$}}
\newcommand{\openn}{\hbox{${\rm I\kern-.2em N}$}}
 
\newcommand{\logit}{\operatorname{logit}}

\newcommand{\CIst}{\textnormal{CI}^{\textnormal{st}}}

\newcommand{\sg}{\textnormal{sg}}
\newcommand{\IF}{\textnormal{IF}}
\newcommand{\Var}{\operatorname{Var}}
\newcommand{\Rem}{\operatorname{Rem}}
  
\usepackage[mathscr]{euscript}
\usepackage{setspace}

\definecolor{dgreen} {RGB}{78,138,21}
\definecolor{gray80}{rgb}{0.8,0.8,0.8}
\definecolor{nred}   {RGB}{224,0,0}
\usepackage{listings}
\title{Evaluating the Impact of Treating the Optimal Subgroup}
\author{Alexander R. Luedtke and Mark J. van der Laan}

 \date{}

\begin{document}
\maketitle

\doublespacing

\begin{abstract}
\singlespacing
Suppose we have a binary treatment used to influence an outcome.
Given data from an observational or controlled study, we wish to
determine whether or not there exists some subset of observed
covariates in which the treatment is more effective than the standard
practice of no treatment. Furthermore, we wish to quantify the improvement in
population mean outcome that will be seen if this subgroup receives
treatment and the rest of the population remains untreated.
We show that this problem is surprisingly challenging given how often it is an (at
least implicit) study objective. Blindly applying standard techniques
fails to yield any apparent asymptotic results, while using existing
techniques to confront the non-regularity does not necessarily help at
distributions where there is no treatment effect.
Here we describe an approach to estimate the impact of treating the
subgroup which benefits from treatment that is valid in a nonparametric
model and is able to deal with the case where there is no treatment effect.
The approach is a slight modification of an approach that recently
appeared in the individualized medicine literature.
\end{abstract}
\newpage

\section{Introduction}
Traditionally, statisticians have evaluated the efficacy of a new treatment using an average treatment effect which compares the population mean outcomes when everyone versus no one is treated. While analyses of marginal effect often successfully identify whether or not introducing a treatment into the population is beneficial, these analyses underestimate the overall benefit of introducing treatment into the population when treatment is on average harmful in some strata of covariates. A treatment need not have adverse physiological side effects for the treatment effect to be negative: it will be negative if the administration of an inferior treatment under study precludes the administration of a superior treatment. To avoid this problem, researchers often perform subgroup analyses to see if the treatment effect varies between different strata of covariates \cite{Assmannetal2000,Rothwell2005}. Many investigators consider subgroups defined by a single covariate at a time \cite{Yusufetal1991,VanderWeele2009}, though there is a growing trend toward defining these subgroups using multiple baseline covariates \cite{Kent2007,Abadie2013}.

Subgroup analyses have led to much disagreement between clinicians and statisticians. As has been highlighted elsewhere \cite{Rothwell2005}, A R Feinstein eloquently described this controversy as follows:
\begin{quote}
``The essence of tragedy has been described as the
destructive collision of two sets of protagonists, both
of whom are correct. The statisticians are right in
denouncing subgroups that are formed post hoc from
exercises in pure data dredging. The clinicians are also
right, however, in insisting that a subgroup is
respectable and worthwhile when established a priori
from pathophysiological principles.'' \cite{Feinstein1998}
\end{quote}
While learning about subgroup specific effects is clearly important when they exist, the concerns of statisticians are understandable. When the analysis is not prespecified, statistical significance procedures tend not to be reliable \cite{Rothwell2005,Yusufetal1991,Lagakos2006}. To exemplify the concerns on this issue, P Sleight shows that the strong marginal effect of aspirin for preventing myocardial infarction changes to a negative effect in two subgroups when these subgroups are defined by astrological sign \cite{Sleight2000}. While it is clearly unlikely in practice that astrological sign yields heterogeneous subgroups, one would hope that a statistical procedure would be sufficiently robust to inform the user that astrological sign is not in fact associated with efficacy.
Some have argued that sample splitting methods would help robustify a procedure to ``data dredging'' by either humans or overfitting by an algorithm \cite{Lipkovichetal2011,Dmitrienkoetal2015,Malanietal2009}.

In a related literature, optimal individualized treatment strategy have been developed to formalize the process of allowing treatment decisions to depend on baseline covariates in a rigorous manner \cite{Chakraborty&Moodie2013}. An individualized treatment strategy is a treatment strategy that makes a treatment decision based on a patient's covariates. Often the objective for such treatments is to optimize the population mean outcome under the given treatment strategy \cite{Murphy03,Robins04}. For binary treatment decisions in a single time point setting, an optimal individualized treatment strategy is any individualized treatment strategy which treats all individuals for which the average treatment effect is positive in their strata of covariates and does not treat anyone for whom the average treatment is negative in their strata of covariates. Estimating the population mean outcome under the optimal individualized treatment rule has been shown to be non-regular when the optimal treatment strategy is not unique \cite{Robins&Rotnitzky14,Luedtke&vanderLaan2015b}. This non-regularity causes standard semiparametric estimation approaches to fail. Despite the complexity of this estimation problem, \cite{Chakrabortyetal2014} shows that one can develop a slower than root-$n$ rate confidence interval for the mean outcome under the estimated optimal individualized treatment rule using a bootstrap procedure, and \cite{Luedtke&vanderLaan2015b} shows how to obtain a root-$n$ rate confidence interval for the actual optimal individualized treatment strategy. Often one can use the same confidence interval for these two estimation problems because one can estimate the optimal treatment strategy consistently (in terms of the strategy's mean outcome) at a faster than root-$n$ rate \cite{Luedtke&vanderLaan2015b}.

As the reader may have noticed, two literatures are analogous -- if one defines the optimal subgroup as the subgroup of covariate strata in which the treatment effect is positive, then the optimal subgroup is (up to covariate strata for which there is no treatment effect) equal to the group of individuals for which an optimal treatment rule suggests treatment. However, the subgroup literature has not confronted the problem of developing high powered inference when an arbitrary algorithm is used to develop the subgroup -- while the sample splitting procedure described in \cite{Malanietal2009} is valid for a single sample split provided the optimal subgroup is not empty, subsequently averaging across sample splits will not yield valid inference (see the discussion of the use of cross-validation for individualized treatment rules in \cite{vanderLaan&Luedtke2014}). Thus there is a significant loss of statistical power in such a procedure.

In this work, we aim to satisfy the desires of both statisticians and clinicians -- we seek a statistically valid subgroup analysis procedure which allows the incorporation of both the subject matter knowledge of physicians and the agnostic flexibility of modern statistical learning techniques. Our subgroup analysis procedure will return an estimate of the population level effect of treating everyone in a stratum of covariates with positive treatment effect versus treating no one. This succinctly characterizes the effect of optimally introducing a given treatment into a population. To estimate this quantity, we modify an estimator from the individualized treatment literature which overcomes a statistical challenge that typically arises when trying to estimate quantities involving individualized treatment rules \cite{Luedtke&vanderLaan2015b}. We will show that an additional statistical challenge arises when trying to use a variant of this estimator in the subgroup setting. We will then show how to overcome this challenge.

\section{Statistical formulation}
Suppose we observe baseline covariates $W$, an indicator of binary treatment $A$, and an outcome $Y$ occuring after treatment and covariates. Let $P_0$ be some distribution for $O\equiv(W,A,Y)$ in a nonparametric statistical model $\mathcal{M}$ that at most places restrictions on the probability of treatment given covariates. We observe $n$ independent individuals $O_1,\ldots,O_n$ drawn from $P_0$. Define $b_0(W)=E_{P_0}[Y|A=1,W]-E_{P_0}[Y|A=0,W]$. Under causal assumptions not elaborated here, $b_0(W)$ can be identified with the additive effect of treatment on outcome if everyone versus no one in a strata of covariates $W$ receives treatment \cite{Robins1986}. We use $\sg$ to denote any (measurable) subset of the support of $W$. Define
\begin{align*}
\Psi_{\sg}(P_0)&\equiv \int_{\sg} b_0(w)dP_0(w).
\end{align*}
Under causal assumptions, $\Psi_{\sg}(P_0)$ is identified with the difference (i)-(ii) between (i) the average outcome if the only individuals receiving treatment in the population are those whose covariates fall in $\sg$ and (ii) the average outcome if no one in the population receives treatment. Drawing parallels to optimal individualized treatment strategies, $\Psi_{\sg}(P_0)$ is maximized at $\sg$ if and only if $\sg$ includes precisely those individuals with covariate $w$ such that $b(w)>0$ and does not include those with $b(w)<0$ \cite{Murphy03,Robins04}. The maximizer is non-unique at so-called ``exceptional laws'', i.e. distributions for which $b(W)=0$ with positive $P_0$ probability \cite{Robins04}. Define $\Psi(P_0)\equiv \max_{\sg}\Psi_{\sg}(P_0)$. Throughout we define $b_P$, $\Psi_{\sg}(P)$, and $\Psi(P)$ at arbitrary $P\in\mathcal{M}$ analogously to $b_0$, $\Psi_{\sg}(P_0)$, and $\Psi(P_0)$.

\section{Breakdown of ``standard'' estimators}
We now describe how the standard semiparametric estimation roadmap seems to suggest we estimate $\Psi(P_0)$. A function known as the efficient influence function (EIF) often plays a key role in this estimation procedure. While we avoid a formal presentation of the derivation of EIFs here, the key result about EIFs is that they typically yield the expansion
\begin{align}
n^{1/2}[\Psi(\hat{P}_n)-\Psi(P_0)]&= -n^{1/2}\int \IF_{\hat{P}_n}(o)dP_0(o) + n^{1/2}\Rem(\hat{P}_n,P_0), \label{eq:nocorrect}
\end{align}
where $\hat{P}_n$ is an estimate of $P_0$, $\IF_{\hat{P}_n}$ is the EIF of $\Psi$ at $\hat{P}_n$, and $\Rem(\hat{P}_n,P_0)$ is a remainder that plausibly converges to zero faster than $n^{-1/2}$. We will present an explicit expression for $\IF_{\hat{P}_n}$ at the end of this section, but for now we state that the corresponding remainder term is given by
\begin{align*}
\Rem(\hat{P}_n,P_0)=&\, \sum_{\tilde{a}=0}^1 (2\tilde{a}-1) \int I(w\in\sg_n)\left[1-\frac{P_0(\tilde{a}|w)}{\hat{P}_n(\tilde{a}|w)}\right]\left(E_{\hat{P}_n}[Y|\tilde{a},w]-E_{P_0}[Y|\tilde{a},w]\right) dP_0(o) \\
&+ \Psi_{\sg_n}(P_0)-\Psi_{\sg_0}(P_0),
\end{align*}
where $\sg_n$ is the optimal subgroup under $\hat{P}_n$. The first term on the right is a double robust term \cite{vdL02} that shrinks to zero faster than $n^{-1/2}$ if the outcome regression and treatment mechanism are estimated well. The second term requires that the optimal subgroup can be estimated well and is plausible if the stratum specific treatment effect function does not concentrate too much mass near zero (mass at zero is not problematic since any subgroup decision for these strata is optimal). See Theorem 8 of \cite{Luedtke&vanderLaan2015b} for precise conditions under which this term is small. In principle $\sg_n$ need not be an optimal subgroup under $\hat{P}_n$, i.e. one can replace $\Psi(\hat{P}_n)$ on the left-hand side of (\ref{eq:nocorrect}) with $\Psi_{\sg_n}(\hat{P}_n)$ without changing the expansion. We ignore such considerations here for brevity, though discussion in a closely related problem is given in \cite{vanderLaan&Luedtke2014}.

A one-step estimator of the form $\psi_n\equiv \Psi(\hat{P}_n)+\frac{1}{n}\sum_{i=1}^n \IF_{\hat{P}_n}(O_i)$ aims to correct the bias on the right-hand side of (\ref{eq:nocorrect}) by adding an estimate of that expectation, yielding
\begin{align}
\sqrt{n}\left[\psi_n-\Psi(P_0)\right]&\approx \frac{1}{\sqrt{n}}\sum_{i=1}^n \left(\IF_{\hat{P}_n}(O_i)-\int\IF_{\hat{P}_n}(o)dP_0(o)\right), \label{eq:firstord}
\end{align}
where the above approximation is valid provided $n^{1/2}\Rem(\hat{P}_n,P_0)$ converges to zero in probability. For a general parameter $\Psi$, targeted minimum loss-based estimators (TMLEs) can be seen to follow the above prescribed formula, with the estimate $\hat{P}_n$ carefully chosen so that the empirical mean of $\IF_{\hat{P}_n}$ is zero, and thus the final estimator is the plug-in estimator $\psi_n=\Psi(\hat{P}_n)$ \cite{vanderLaan&Rose11}. A detailed exposition of efficiency theory is given in \cite{Bickel1993}.


We cannot apply the central limit theorem to the right-hand side of (\ref{eq:firstord}) without further conditions because the right-hand side is a root-$n$ empirical mean over functions which depend on the data. We now give sufficient conditions. Suppose that $\IF_{\hat{P}_n}$ has a limit $\IF_{\infty}$ in the sense that
\begin{align}
\left(\IF_{\hat{P}_n}-\IF_{\infty}\right)^2\textnormal{ has $P_0$ expectation converging to zero.} \tag{Lim} \label{eq:l2}
\end{align}
Typically $\IF_\infty=\IF_{P_0}$. If $\hat{P}_n$ is not allowed to heavily overfit the data, the instances of $\IF_{\hat{P}_n}$ on the right-hand side of (\ref{eq:firstord}) can be replaced with $\IF_{\infty}$. The conditions on $\hat{P}_n$ which prevent overfitting are given by the empirical process conditions presented in Part 2 of \cite{vanderVaartWellner1996}. If
\begin{align}
\Var_{P_0}[\IF_{\infty}(O)]>0, \tag{V+} \label{eq:posvar}
\end{align}
then the central limit theorem can be used to see that $n^{1/2}[\psi_n-\Psi(P_0)]$ converges to a normal distribution with mean zero and variance $\Var_{P_0}[\IF_{\infty}(O)]$. Under these conditions, $\Psi(P_0)$ falls in $\psi_n\pm 1.96\frac{\sigma_n}{\sqrt{n}}$ with probability approaching 0.95, where $\sigma_n^2$ is the empirical variance of $\IF_{\hat{P}_n}$ applied to the data. If $\Var_{P_0}[\IF_{\infty}(O)]$ is zero, then $n^{1/2}[\psi_n-\Psi(P_0)]$ converges to zero in probability, but there is no guarantee that $\psi_n\pm 1.96\frac{\sigma_n}{\sqrt{n}}$ contains $\Psi(P_0)$ with probability approaching 0.95: both $\sqrt{n}[\psi_n-\Psi(P_0)]$ and $\sigma_n$ are converging to zero, but coverage depends on the relative rate of convergence of the two quantities.

We now argue that it is unlikely that both (\ref{eq:l2}) and (\ref{eq:posvar}) hold. When $\hat{P}_n$ is non-exceptional,
\begin{align*}
\IF_{\hat{P}_n}(o)&= I[b_{\hat{P}_n}(w)>0]\left[\frac{2a-1}{\hat{P}_n(a|w)}(Y-E_{\hat{P}_n}[Y|A=a,W=w]) + b_{\hat{P}_n}(w)\right] - \Psi({\hat{P}_n}). 
\end{align*}
If $\hat{P}_n$ is exceptional, the above definition of $\IF_{\hat{P}_n}$ can still be used and the same central limit theorem result about (\ref{eq:firstord}) holds under (\ref{eq:l2}) and (\ref{eq:posvar}), though in truth $\Psi$ is not smooth enough at $\hat{P}_n$ for an efficient influence function to be well-defined \cite{Robins&Rotnitzky14,Luedtke&vanderLaan2015b}. In light of the above expression, the validity of (\ref{eq:l2}) will typically require $I[b_{\hat{P}_n}(W)>0]$ to have a mean-square limit. Suppose the data is drawn from an exceptional law where the treatment effect is zero on some set $S_0$. In that case we do not expect $I[b_{\hat{P}_n}(W)>0]$ to converge to anything on $S_0$ since likely $b_{\hat{P}_n}(w)$ does not converge to 0 strictly from above or below at any given $w$ for which $b_0(w)=0$.

Now consider the case where treatment is always harmful, i.e. $b_0(W)<0$ with probability 1. In this case $b_0(W)\ge 0$ with probability 0, and so we would expect the indicator that $b_{\hat{P}_n}(w)$ is positive converges to zero if $\hat{P}_n$ is a good estimate of $P_0$. But in this case the subgroup that should be treated is empty so that if the limit $\IF_{\infty}$ exists then it is zero almost surely and (\ref{eq:posvar}) does not hold.

Finally, consider the intermediate case where there is no additive treatment effect within any strata of covariates, i.e. $b_0(w)=0$ for all $w$. If on any positive probability set convergence occurs from both above and below, then we do not expect (\ref{eq:l2}) to hold. If $b_{\hat{P}_n}(w)$ converges to zero from below for all $w$, then we expect (\ref{eq:l2}) to hold with $\IF_\infty$ equal to the constant function zero and (\ref{eq:posvar}) not to hold. If, for each $w$, $b_{\hat{P}_n}(w)$ converges to zero from either strictly above or strictly below and the set of covariates for which the convergence occurs from above happens with positive probability, then we expect (\ref{eq:l2}) and (\ref{eq:posvar}) to hold.

\section{Avoiding the need for (\ref{eq:l2}) and (\ref{eq:posvar})}
In this section, we present an estimator which overcomes both (\ref{eq:l2}) and (\ref{eq:posvar}). We first present an estimator which does not require (\ref{eq:l2}), and then argue that a simple extension of this estimator also does not require (\ref{eq:posvar}).

This estimation strategy is similar to the one-step estimator presented in the previous section, but designed to estimate the parameter in an online fashion which eliminates the need for convergence. The online one-step estimator was originally presented in \cite{vanderLaan&Lendle2014}, and was refined in \cite{Luedtke&vanderLaan2015b} to deal with cases where the convergence of the sort required by (\ref{eq:l2}) fails to hold. The method will be presented in full generality in a forthcoming paper.

Let $\hat{P}_n^i$ represent an estimate of $P_0$ based on observations $O_1,\ldots,O_i$. The stabilized online one-step estimator for $\Psi(P_0)$ is given by
\begin{align}
\psi_n^{\textnormal{st}}&\equiv \frac{\bar{\sigma}_n}{n-\ell_n}\sum_{i=\ell_n}^{n-1}\frac{\Psi(\hat{P}_n^i) + \IF_{\hat{P}_n^i}(O_{i+1})}{\hat{\sigma}_i}. \label{eq:psinstab}
\end{align}
where $\ell_n$ is some user-defined quantity that may or may not grow to infinity but must satisfy $n-\ell_n\rightarrow\infty$, $\hat{\sigma}_i^2$ represents an estimate of the variance of $\IF_{\hat{P}_n^i}^{d_i}(O)$ based on observations $O_1,\ldots,O_i$, and $\bar{\sigma}_n$ is the harmonic mean $1/[\frac{1}{n-\ell_n}\sum_{i=\ell_n}^{n-1} \hat{\sigma}_i^{-1}]$ of $\hat{\sigma}_{\ell_n},\ldots,\hat{\sigma}_{n-1}$. We now wish to apply the martingale central limit theorem \cite{Brown1971} to understand the behavior of $\sqrt{n-\ell_n}\bar{\sigma}_n^{-1}[\psi_n^{\textnormal{st}}-\Psi(P_0)]$. As each term in the sum defining $\psi_n$ has variance converging to 1 due to the stabilization by $\hat{\sigma}_i$, the validity of our central limit theorem argument does not rely on an analogue of (\ref{eq:l2}). It does, however, rely on an analogue of (\ref{eq:posvar}). The primary condition we would use to establish the validity of the central limit theorem argument is that the set of $\hat{\sigma}_i^2$ are consistent for $\Var_{P_0}\left[\IF_{\hat{P}_n^i}(O_{i+1})\right]$ as $i$ gets large and that there exists some $\delta>0$ such that
\begin{align*}
\Var_{P_0}\left[\IF_{\hat{P}_n^i}(O_{i+1})\right]>\delta^2>0\textnormal{ for all $i$ with probability approaching 1.} \tag{V+'} \label{eq:posvarp}
\end{align*}
The former condition holds under a Glivenko-Cantelli condition which is discussed in Theorem 7 of \cite{Luedtke&vanderLaan2015b}. Under these conditions, Section 7 of \cite{Luedtke&vanderLaan2015b} (especially Lemma 6) shows that
\begin{align*}
\frac{\sqrt{n-\ell_n}\left[\psi_n^{\textnormal{st}}-\Psi(P_0)\right]}{\bar{\sigma}_n}&\approx \frac{1}{\sqrt{n-\ell_n}}\sum_{i=\ell_n}^{n-1}\frac{\IF_{\hat{P}_n^i}(O_{i+1})-\int \IF_{\hat{P}_n^i}(o) dP_0(o)}{\hat{\sigma}_i}
\end{align*}
provided the same conditions needed for (\ref{eq:firstord}) hold. The above approximation is accurate up to a term that goes to zero in probability. The martingale central limit theorem can now be applied to establish the validity of the 95\% confidence interval $\CIst\equiv \left[\psi_n^{\textnormal{st}}\pm 1.96\frac{\bar{\sigma}_n}{\sqrt{n-\ell_n}}\right]$. We refer the reader to Theorem 2 in \cite{Luedtke&vanderLaan2015b} for a sense of the formal conditions needed to prove this result. If the treatment effect is negative for all strata of covariates, then (\ref{eq:posvarp}) will not hold if $\hat{P}_n^i$ is a reasonable estimate of $P_0$ in the sense that the estimated optimal subgroup converges to the empty set. Similarly, we have no guarantee that (\ref{eq:posvarp}) will hold if $b_0(W)$ is zero almost surely.

Suppose that we are not willing to assume that (\ref{eq:posvarp}) holds. A natural approach is to redefine the inverse weights to equal $\hat{\sigma}_i(\delta)\equiv \hat{\sigma}_i\vee \delta$ for some fixed $\delta>0$. We then define $\bar{\sigma}_n(\delta)$ to be equal to the harmonic mean of $\hat{\sigma}_{\ell_n}(\delta),\ldots,\hat{\sigma}_{n-1}(\delta)$ and $\psi_n^{\textnormal{st}}(\delta)$ as in (\ref{eq:psinstab}) but with $\hat{\sigma}_i$ and $\bar{\sigma}_n$ replaced by $\hat{\sigma}_i(\delta)$ and $\bar{\sigma}_n(\delta)$. Clearly $\bar{\sigma}_n(\delta)\ge \bar{\sigma}_n$, and thus the confidence interval $\CIst(\delta)\equiv \left[\psi_n^{\textnormal{st}}(\delta)\pm 1.96\frac{\bar{\sigma}_n(\delta)}{\sqrt{n-\ell_n}}\right]$ is no wider than that $\CIst$, though it may have a different midpoint. We conjecture that this confidence interval is conservative when the truncation scheme is active so that $\bar{\sigma}_n(\delta)>\bar{\sigma}_n$, though proving this result has proven challenging.

To give the reader a sense of why we have hope that this conjecture will hold, we show in the appendix that adding normal noise (with random variance depending on the sample) to $\psi_n(\delta)$ can yield a valid 95\% confidence interval for $\Psi(P_0)$. It then seems reasonable that removing the noise can only improve coverage. Readers whose primary concern is the theoretical soundness of an inferential procedure can apply the estimator in the appendix and rest assured that their confidence interval will be valid provided the optimal subgroup is estimated sufficiently well and a double robust term is small. Nonetheless, the type I error gains by not using this noised estimator, which we will see provided the conjecture holds, would seem to imply that our original unnoised confidence interval performs better than the noised interval.

\section{Simulation study}
\subsection{Methods}
We now present a simulation study conducted in \texttt{R} \cite{R2014}. Our simulation uses a four-dimensional covariate $W$ drawn from a mean zero normal distribution with identity covariance matrix. Treatment $A$ is drawn according to a Bernoulli random with probability of success $1/2$, independent of baseline covariates. The outcome $Y$ is Bernoulli, and the outcome regressions considered in our primary analysis are displayed in Table \ref{tab:sims}.

\begin{table}[h!]
\begin{tabular}{l l l l l}
Simulation & $\logit E[Y|A,W]$ & $\Psi(P_0)$ & $P_0\left(b_0(W)=0\right)$ & $P_0\left(b_0(W)<0\right)$ \\
\hline
N1 & $W_1 + W_2$ & 0 & 1 & 0 \\
N2 & $-0.2A[(W_1-z_{0.8})^+]^2$ & 0 & 0.80 & 0.20 \\
N3 & $-0.25A$ & 0 & 0 & 1 \\
A1 & $0.8A$ & 0.19 & 0 & 0 \\
A2 & $AW_1^+ - AW_2^+$ & 0.06 & 0.25 & 0.38 \\
A3 & $AW_1$ & 0.09 & 0 & 0.50
\end{tabular}
\caption{Data generating distributions for simulation. Decimals rounded to the nearest hundredth. For N2, $z_{0.8}\approx 0.84$ is the 80th percentile of a standard normal distribution. We use $x^+$ to denote the positive part of a real number $x$.}
\label{tab:sims}
\end{table}


We compare two estimators of $\Psi(P_0)$. The first is the stabilized one-step estimator. We truncate the inverse weights at $0.1$, $0.001$, and $10^{-20}$. The results were essentially identical for truncations of $10^{-3}$ and $10^{-20}$, and thus we only display the results for the truncation of $10^{-3}$. We set $\ell_n=n/10$, and to speed up the computation time we estimated the subgroup and outcome regression using observations $O_1,\ldots,O_{k(i)n/10}$ for all $i\ge \ell_n$, where $k(i)$ is the largest integer such that $k(i)n/10<i$ (see Section 6.1 of \cite{Luedtke&vanderLaan2015b} for more details). The second estimator is a ten-fold cross-validated TMLE (CV-TMLE). This estimator is analogous the CV-TMLE for the mean outcome under an optimal treatment rule as presented in \cite{vanderLaan&Luedtke2014}, but is modified to account for the fact that $\Psi(P_0)$ is equal to this quantity minus the mean outcome when no one in the population is treated. We truncate the variance estimates for this estimator at the same values as considered for the stabilized one-step estimator. Following the theoretical results in \cite{vanderLaan&Luedtke14b}, we can formally show that this estimator is asymptotically valid when cross-validated analogues of (\ref{eq:l2}) and (\ref{eq:posvar}) hold. 

We estimate the blip function using the super-learner methodology as described in \cite{Luedtke&vanderLaan2014}. Super-learner is an ensemble algorithms with an oracle guarantee ensuring that the resulting blip function estimate will perform at least as well as the best candidate in the library up to a small error term. We use a squared error loss to estimate the blip function, and use as candidate algorithms \texttt{SL.gam}, \texttt{SL.glm}, \texttt{SL.glm.interaction}, \texttt{SL.mean}, and \texttt{SL.rpart} in the \texttt{R} package \texttt{SuperLearner} \cite{SuperLearner2013}. The outcome regression $E[Y|A,W]$ is estimated using this same super-learner library but with the log-likelihood loss to respect the bounds on the outcome. The probability of treatment given covariates was treated as known and the known value was used by all of the estimators.

We also compare our estimators to the oracle estimator which \textit{a priori} knows the optimal subgroup. In particular, we use a CV-TMLE for $\Psi_{\sg_0}(P_0)$ in which we treat $\sg_0$ as known. The estimation problem is regular in this case so that we expect the corresponding confidence intervals to have proper coverage at exceptional laws. We truncate the variance estimate at the same values as for the other methods.

\subsection{Results}
\begin{figure}[h!]
  \centering
  \includegraphics[width=\textwidth]{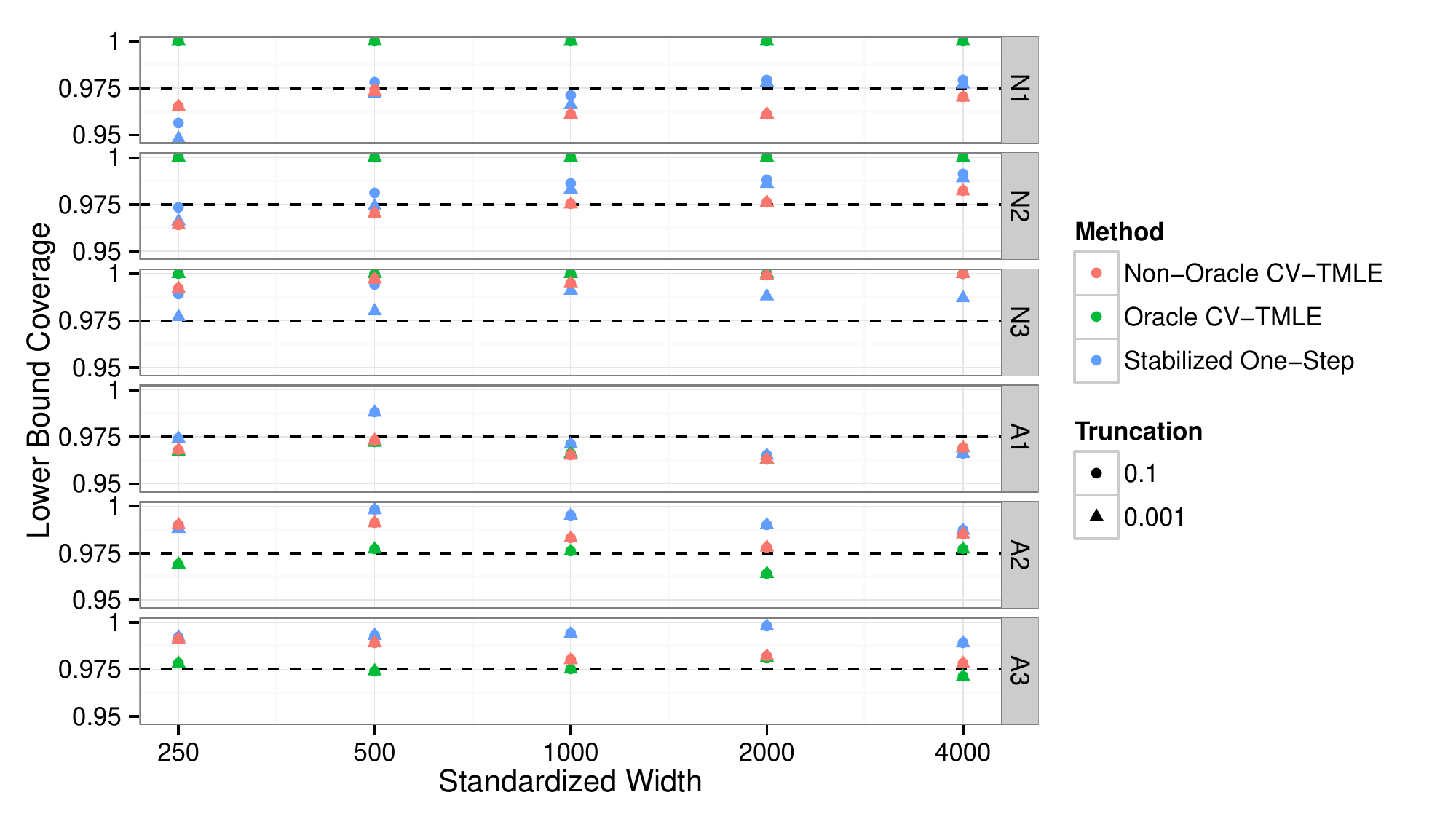}
  \caption{Coverage of 97.5\% lower confidence intervals for the impact of treating the optimal subgroup versus sample size. Horizontal axis on a log scale. All methods (approximately) achieve nominal coverage, with the stabilized one-step and non-oracle CV-TMLE generally being conservative.}
  \label{fig:lb}
\end{figure}

Figure \ref{fig:lb} displays the coverage of the confidence interval lower bounds of the various estimation strategies. All methods appear to achieve proper 97.5\% lower bound coverage, with the stabilized one-step and the non-oracle CV-TMLE estimators generally being conservative. This conservative behavior is to be expected given that both of these sample splitting procedures need to estimate the optimal subgroup, and thus in any finite sample are expected to be negatively biased due to the resulting suboptimal subgroup used from the estimate. The oracle CV-TMLE attains the nominal coverage rate for alternative distributions, and for non-alternatives is also conservative.

\begin{figure}[h!]
  \centering
  \includegraphics[width=\textwidth]{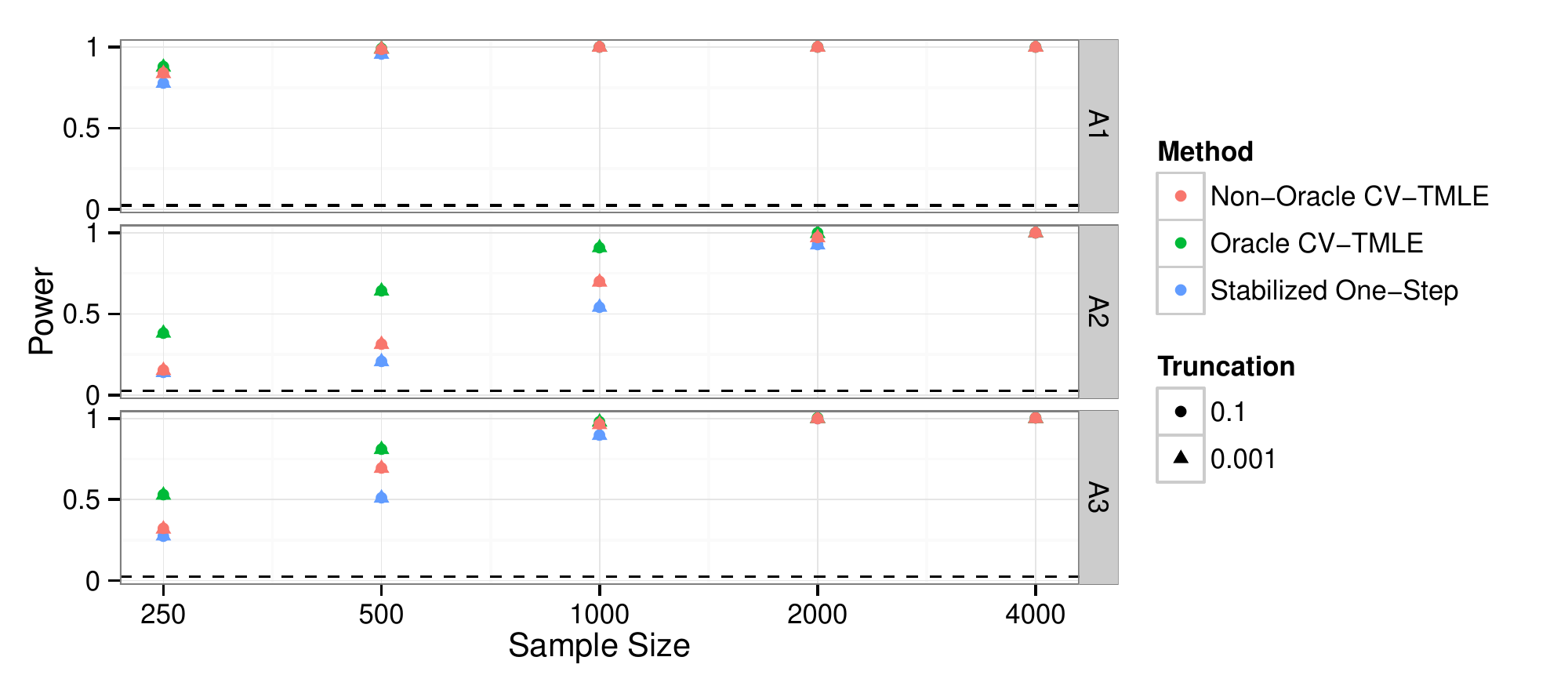}
  \caption{Power for rejecting a test of $H_0 : \Psi(P_0)=0$ versus $H_1 : \Psi(P_0)>0$ at the 0.025 level at different sample sizes. Horizontal axis on a log scale. All methods have power increasing to 1. The stabilized one-step has lower power than the non-oracle CV-TMLE, though its power is typically more comparable to the stabilized one-step relative than the oracle CV-TMLE which knew the optimal subgroup from the outset.}
  \label{fig:pow}
\end{figure}

We now verify the tightness of the lower bounds for the alternative distributions A1, A2, and A3. Figure \ref{fig:pow} shows the power for the test of $H_0 : \Psi(P_0)=0$ against $H_1 : \Psi(P_0)>0$, where the test was conducted using the duality between hypothesis tests and confidence intervals. We see that the stabilized one-step is slightly less powerful than the non-oracle CV-TMLE. This is likely due to the online nature of the stabilized one-step estimator relative to the CV-TMLE. Nonetheless, the power loss is not large and the fact that we have actual theoretical results for this estimator even at exceptional laws should make up for this slight loss of power.

\begin{figure}[h!]
  \centering
  \includegraphics[width=\textwidth]{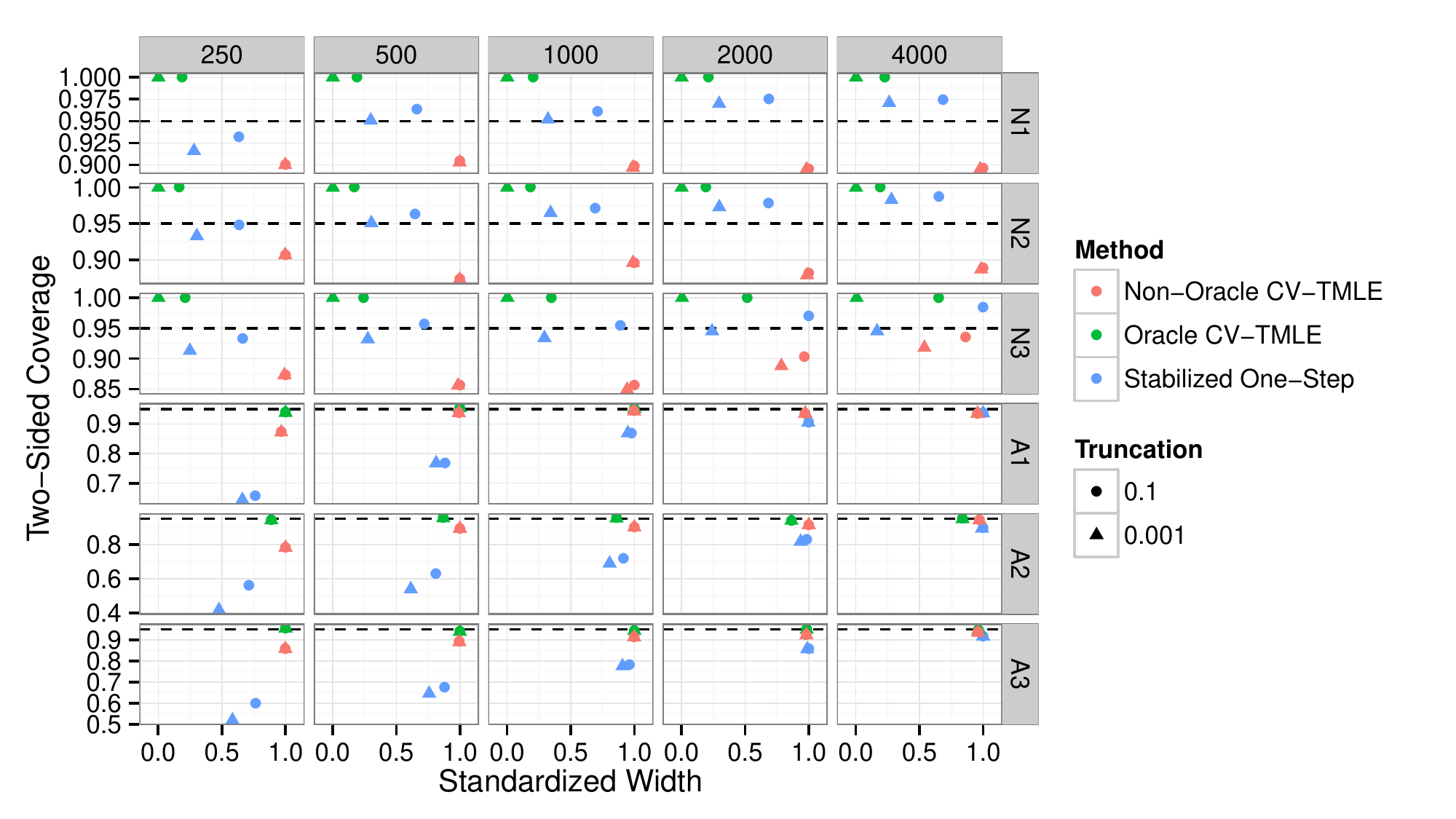}
  \caption{Coverage of 95\% two-sided confidence intervals for the impact of treating the optimal subgroup versus the average confidence interval width (standardized so that the maximum width in a given sample size-data generating distribution pair is 1). The stabilized one-step confidence intervals have nominal coverage for all null distributions, and have coverage approaching nominal for all alternative distributions. The non-oracle CV-TMLE achieves near nominal coverage for all alternative distributions, though has below nominal coverage of approximately 90\% for all of the null distributions.}
  \label{fig:covg}
\end{figure}

Figure \ref{fig:covg} displays the two-sided coverage of the 95\% confidence intervals. One could argue that upper bound coverage is not interesting for the null distributions, given that any failure of the upper bound of the confidence interval to cover $\Psi(P_0)=0$ requires this upper bound to be negative. Hence we can always obtain proper upper bound coverage at null distributions by ensuring that the upper bound of our confidence interval respects the parameter space of $\Psi$, i.e. is nonnegative. Nonetheless, the coverage of the uncorrected two-sided confidence intervals (upper bound may be negative) is useful for detecting a lack of asymptotic normality of the estimator sequence. While the stabilized one-step has two-sided coverage above 0.95 for all distributions at all sample sizes of at least 500, the coverage for the unadjusted non-oracle CV-TMLE confidence interval falls at or below 0.90 for N1 and N2 at all sample sizes. This is in line with our lack of asymptotic results for the non-oracle CV-TMLE at exceptional laws.

We now consider the two-sided coverage for the alternative distributions A1, A2, and A3. The stabilized one-step confidence intervals have coverage that improves with sample size, though the improvement appears slow. The coverage is near nominal at a sample size of 4000 for all three simulations. In light of Figure \ref{fig:lb}, essentially all of the coverage deficiency is a result of a failure of the upper bound. This makes sense given that our estimator relies on a second-order term measuring a linear combination of the difference in impact of treating the estimated subgroups (estimated on increasing chunks of data) versus treating the optimal subgroup. While this term often reasonably shrinks to zero faster than $n^{-1/2}$, in finite samples this term can hurt the upper bound coverage. The non-oracle CV-TMLE confidence intervals, on the other hand, attain near nominal coverage at large sample sizes. This is to be expected at the non-exceptional laws A1 and A3 given that we can prove asymptotic normality in this case. We do not have an asymptotic result supporting the method's proper coverage for exceptional law A2, though this is an interesting area for future work.



\section{Discussion}
We have studied the statistical challenges associated with estimating the additive effect of treating the subgroup of individuals with positive stratum specific additive treatment effect. We showed that these challenges are similar to those arising when estimating the mean outcome under an optimal individualized treatment strategy. Indeed, the individuals treated by the strategy which maximizes the population mean outcome are the same individuals who belong to the optimal subgroup. An additional challenge arises when one wishes to consider the relative measure giving the additive effect of treating only those individuals in the optimal subgroup versus treating no one in the population. In this case the parameter of interest is estimable at a faster than root-$n$ rate for some data generating distributions. Procedures which yield root-$n$ rate confidence intervals tend to fail in this setting due to the need to estimate both the (in truth empty) optimal subgroup and the variance of the estimate of the impact of treating this subgroup: generally the subgroup estimate will converge to the empty set and the variance estimate will converge to zero, but there is no guarantee that the relative rate of convergence of the two will yield valid inference.

Despite this added inferential challenge, we argue that obtaining a confidence interval for the impact of treating the optimal subgroup requires only minor modification to the confidence interval for the mean outcome when only the optimal subgroup is treated. In particular, we propose truncating the estimated variance in the martingale sum used in \cite{Luedtke&vanderLaan2015b} at some constant $\delta>0$. If the truncation is not active, which will typically be true for alternative distributions under which there exists a subgroup for which the treatment effect is positive and is arguably true for many null distributions as well, then, under standard regularity conditions, we obtain root-$n$ rate inference with coverage approaching 0.95. If the data is generating according to an alternative distribution for which there is a non-null (positive or negative) treatment effect within all strata of covariates, then our estimator is asymptotically efficient and, provided the truncation is not active, our confidence interval is asymptotically equivalent to a standard Wald-type confidence interval (see Corollary 3 in \cite{Luedtke&vanderLaan2015b}). We expect our confidence interval to be conservative when the truncation is active, though we leave this as a conjecture. We have instead shown that adding noise to our estimator yields a confidence interval with proper 95\% coverage, though we suggest using the unnoised estimator in practice. 

One could imagine several alternative solutions to the described inferential challenge. One such solution is to ensure that the variance of our estimator minus the truth, scaled by root-$n$, is positive as sample size grows. This can be accomplished by changing the definition of the optimal subgroup to ensure that this subgroup is not too small, e.g. contains at least 10\% of the population. One can show via a change of variables that estimating the mean outcome under such a constrained subgroup is equivalent to estimating the mean outcome under an optimal rule which can treat at most 90\% of the population, see \cite{Luedtke&vanderLaan2015}. Estimating this alternative constrained parameter is still difficult when the optimal subgroup is non-unique, though there is little risk of degenerate first-order behavior in this case. To construct confidence intervals despite the non-uniqueness of the optimal subgroup, one can combine the results in \cite{Luedtke&vanderLaan2015} with the stabilized one-step estimator presented in \cite{Luedtke&vanderLaan2015b}.

A cross-validated TMLE, closely related to that presented in \cite{vanderLaan&Luedtke14b}, outperformed the method proposed in this paper in many simulation settings. Nonetheless, we do not have any asymptotic results about the CV-TMLE at exceptional laws, in contrast to the estimator presented in this paper for which we do have such results. This estimator's lack of asymptotic normality at such laws was evident in our simulation. We view a careful study of this estimator's behavior at exceptional laws to be an important area for future research. In a forthcoming work, we will present a stabilized TMLE that has the same desirable asymptotic properties of the stabilized one-step estimator but, like the CV-TMLE, is a substitution estimator (thereby forcing the estimate to respect the parameter space).

One could imagine considering other parameters relating to the optimal subgroup that we have presented in this paper. For example, investigators may be interested in estimating the impact of treating everyone in the optimal subgroup on some secondary outcome $\tilde{Y}$. Each such parameter yields a new estimation problem and, in our experience, many of these problems still face at least one of the two primary challenges that we faced in this paper. In particular, these problems are often non-regular when the optimal subgroup is non-unique, and may have degenerate first-order behavior when the optimal subgroup is empty.

\section*{Acknowledgement}
This research was supported by NIH grant R01 AI074345-06. Alex Luedtke was supported by a Berkeley Fellowship. The authors thank Tyler VanderWeele for valuable discussions.

{\singlespacing
\small
\bibliographystyle{vancouver}
 \bibliography{persrule}}

\section*{Appendix}
Let $\{Z_i : i=1,\ldots,\infty\}$ be a sequence of i.i.d. normal random variable independent of all other sources of randomness under consideration. Let
\begin{align*}
\tilde{\psi}_n^{\textnormal{st}}(\delta)&= \frac{\bar{\sigma}_n}{n-\ell_n}\sum_{i=\ell_n}^{n-1}\frac{\Psi(\hat{P}_n^i) + \IF_{\hat{P}_n^i}(O_{i+1}) + \sqrt{(\delta^2-\hat{\sigma}_i^2)^+}Z_i}{\hat{\sigma}_i(\delta)},
\end{align*}
where $x^+$ is the positive part of a real number $x$. Observe that each term in the sum on the right-hand side above has variance approximately $1$ (approximately because $\hat{\sigma}_i$ is only an estimate of $\Var_{P_0}[\IF_{\hat{P}_n^i}(O)]$). It will then follow that, under regularity conditions, we can apply the martingale central limit theorem appearing in \cite{Brown1971} to show that $\left[\tilde{\psi}_n^{\textnormal{st}}(\delta)\pm 1.96\frac{\bar{\sigma}_n(\delta)}{\sqrt{n-\ell_n}}\right]$ has coverage approaching 95\% for $\Psi(P_0)$. These regularity conditions are the same as those used to establish the validity of $\CIst$, except that they do not require (\ref{eq:posvarp}) to hold. Note the similarity to $\CIst(\delta)$, though the above confidence interval is centered about $\tilde{\psi}_n(\delta)$ rather than $\psi_n(\delta)$.

We now relate the noised $\tilde{\psi}_n^{\textnormal{st}}$ to the unnoised $\psi_n^{\textnormal{st}}$. Conditional on the data, $\tilde{\psi}_n^{\textnormal{st}}(\delta)$ is equal in distribution to $\psi_n^{\textnormal{st}}(\delta) + \frac{\tilde{\sigma}_n(\delta)}{\sqrt{n-\ell_n}}Z_1$, where $\tilde{\sigma}_n^2(\delta)\equiv \frac{1}{n-\ell_n}\sum_{i=\ell_n}^{n-1} \frac{(\delta^2-\sigma_i^2)^+}{\hat{\sigma}_i^2\vee \delta^2}$. That is, $\tilde{\psi}_n^{\textnormal{st}}(\delta)$ is equal to $\psi_n(\delta)$ plus normal noise, where the variance of the normal noise depends on the data. If the truncation is not active, then the variance of this noise is zero. Otherwise, the variance is positive, but the sign of the noise is independent of the data. Thus it seems reasonable to expect that the unnoised $\psi_n^{\textnormal{st}}$ provides a better estimate of $\Psi(P_0)$ than $\tilde{\psi}_n^{\textnormal{st}}$. It is for this reason that we expect the unnoised confidence interval $\CIst(\delta)$ to have coverage at least 0.95 in large samples.

\end{document}